\documentclass[aps,a4paper,superscriptaddress,showpacs,preprintnumbers,amsmath,amssymb]{revtex4}

\usepackage{ulem}

\usepackage{psfrag} \usepackage{graphicx} \usepackage{dcolumn}
\usepackage{color} \usepackage{latexsym,amsfonts} \usepackage{bm}
\usepackage{amssymb}
\begin{document}

\title{Precision analysis \\ of pseudoscalar interactions in neutron
  beta decays}

\author{A. N. Ivanov}\email{ivanov@kph.tuwien.ac.at}
\affiliation{Atominstitut, Technische Universit\"at Wien, Stadionallee
  2, A-1020 Wien, Austria}
\author{R. H\"ollwieser}\email{roman.hoellwieser@gmail.com}
\affiliation{Atominstitut, Technische Universit\"at Wien, Stadionallee
  2, A-1020 Wien, Austria}\affiliation{Department of Physics,
  Bergische Universit\"at Wuppertal, Gaussstr. 20, D-42119 Wuppertal,
  Germany} \author{N. I. Troitskaya}\email{natroitskaya@yandex.ru}
\affiliation{Atominstitut, Technische Universit\"at Wien, Stadionallee
  2, A-1020 Wien, Austria}
\author{M. Wellenzohn}\email{max.wellenzohn@gmail.com}
\affiliation{Atominstitut, Technische Universit\"at Wien, Stadionallee
  2, A-1020 Wien, Austria} \affiliation{FH Campus Wien, University of
  Applied Sciences, Favoritenstra\ss e 226, 1100 Wien, Austria}
\author{Ya. A. Berdnikov}\email{berdnikov@spbstu.ru}\affiliation{Peter
  the Great St. Petersburg Polytechnic University, Polytechnicheskaya
  29, 195251, Russian Federation}

\date{\today}

\begin{abstract}
We analyze the contributions of the one--pion--pole (OPP) exchange,
caused by strong low--energy interactions, and the pseudoscalar
interaction beyond the Standard Model (BSM) to the correlation
coefficients of the neutron $\beta^-$--decays for polarized neutrons,
polarized electrons and unpolarized protons. The strength of
contributions of pseudoscalar interactions is defined by the effective
coupling constant $C_{ps} = C^{(\rm OPP)}_{ps} + C^{(\rm BSM)}_{ps}$.
We show that the contribution of the OPP exchange is of order $C^{(\rm
  OPP)}_{ps} \sim - 10^{-5}$. The effective coupling constant $C^{(\rm
  BSM)}_{ps}$ of the pseudoscalar interaction BSM can be in principle
complex. Using the results, obtained by Gonza\'lez-Alonso {\it et
  al.}( Prog. Part. Nucl. Phys. {\bf 104}, 165 (2019)) we find that
the values of the real and imaginary parts of the effective coupling
constant $ C^{(\rm BSM)}_{ps}$ are constrained by $ - 3.5 \times
10^{-5} < {\rm Re}\,C^{(\rm BSM)}_{ps} < 0$ and ${\rm Im}\,C^{(\rm
  BSM)}_{ps} < - 2.3 \times 10^{-5}$, respectively. The obtained
results can be used as a theoretical background for experimental
searches of contributions of interactions BSM in asymmetries of the
neutron $\beta^-$--decays with a polarized neutron, a polarized
electron and an unpolarized proton at the level of accuracy of a few
parts of $10^{-5}$ or even better (Abele, Hyperfine Interact. {\bf
  237}, 155 (2016)).
\end{abstract}
 \maketitle

\section{Introduction}
\label{sec:introduction}

Nowadays the neutron lifetime and correlation coefficients of the
neutron $\beta^-$-decays for polarized neutrons, polarized electrons
and unpolarized protons are calculated within the Standard Model (SM)
at the level of $10^{-3}$ including the radiative corrections of order
$O(\alpha/\pi)$ of and corrections caused by the weak magnetism and
proton recoil of order $O(E_e/M)$
\cite{Bilenky1959}--\cite{Ivanov2019a}, where $\alpha$, $E_e$ and $M$
are the fine--structure constant \cite{PDG2018}, an electron energy
and the nucleon mass, respectively. Such a SM theoretical background
has allowed to make steps forwards investigations of contributions of
interactions beyond the SM (BSM) of order $10^{-4}$ or even smaller
\cite{Abele2016}. The analysis of interactions beyond the $V - A$
effective theory of weak interactions \cite{Feynman1958,
  Sudarshan1958, Marshak1959, Nambu1960, Marshak1969} (see also
\cite{Shekhter1959a, Shekhter1959b}) in the neutron $\beta^-$--decays
with different polarizations of massive fermions has a long history
and started in 50th of the 20th century and is continuing at present
time \cite{Lee1956}--\cite{Severijns2019} (see also \cite{Gudkov2006,
  Ivanov2013, Ivanov2017d}). The most general form of the Lagrangian
of interactions BSM has been written in
\cite{Lee1956}-\cite{Severijns2006}, including non--derivative vector
$\bar{\psi}_p\gamma_{\mu}\psi_n$, axial--vector
$\bar{\psi}_p\gamma_{\mu}\gamma^5\psi_n$, scalar $\bar{\psi}_p\psi_n$,
pseudoscalar $\bar{\psi}_p\gamma^5 \psi_n$ and tensor
$\bar{\psi}_p\sigma_{\mu\nu} \psi_n$ nucleon currents coupled to
corresponding lepton currents in the form of local nucleon--lepton
current--current interactions, where $\{1, \gamma_{\mu},
\gamma_{\mu}\gamma^5, \gamma^5, \sigma_{\mu\nu}\}$ are the Dirac
matrices \cite{Itzykson1980}, With respect to $G$--parity
transformations \cite{Lee1956a}, i.e. $G = C\,e^{\,i \pi I_2}$, where
$C$ and $I_2$ are the charge conjugation and isospin operators
\cite{Itzykson1980}, the vector, axial--vector, pseudoscalar and
tensor nucleon currents are $G$--even and the scalar nucleon current
is $G$--odd.  According to the $G$--transformation properties of
hadronic currents, Weinberg divided hadronic currents into two
classes, which are $G$--even first class and $G$--odd second class
currents \cite{Weinberg1958}, respectively. Thus, following Weinberg's
classification the non--derivative vector, axial--vector, pseudoscalar
and tensor nucleon currents in the interactions BSM, introduced in
\cite{Lee1956}--\cite{Severijns2006}, are the first class currents,
whereas the non--derivative scalar nucleon current is the second class
one (see also \cite{Ivanov2018}).  The analysis of superallowed $0^+
\to 0^+$ nuclear beta transitions by Hardy and Towner \cite{Hardy2015}
and Gonz\'alez--Alonso {\it et al.}  \cite{Severijns2019} has shown
that the phenomenological coupling constants of non--derivative scalar
current--current nucleon--lepton interaction is of order $10^{-5}$ or
even smaller. This agrees well with estimates of contributions of the
second class currents, caused by derivative scalar
$\partial^{\mu}(\bar{\psi}_p\psi_n)$ and pseudotensor
$\partial^{\nu}(\bar{\psi}_p\sigma_{\mu\nu}\gamma^5 \psi_n)$ nucleon
currents proposed by Weinberg \cite{Weinberg1958}, to the neutron
lifetime and correlation coefficients of the neutron $\beta^-$--decays
carried out by Gardner and Plaster \cite{Gardner2001, Gardner2013} and
Ivanov {\it et al.}  \cite{Ivanov2017d, Ivanov2019}. The contemporary
experimental sensitivities $10^{-4}$ or even better \cite{Abele2016}
of experimental analyses of parameters of neutron $\beta^-$--decays
(see, for example, \cite{Abele2018, Seng2018, Seng2018a}) demand a
theoretical background for the neutron lifetime and correlation
coefficients of the neutron $\beta^-$--decays with different
polarizations of massive fermions at the level of $10^{-5}$
\cite{Ivanov2017b, Ivanov2017d, Ivanov2019,Ivanov2019a}. As has been
shown in \cite{Cirigliano2010}--\cite{Cirigliano2013a} in the linear
approximation the contributions of vector and axial--vector
interactions BSM can be absorbed by the matrix element $V_{ud}$ of the
Cabibbo--Kobayashi--Maskawa (CKM) mixing matrix and by the axial
coupling constant $\lambda$ (see also \cite{Ivanov2013, Ivanov2017b,
  Ivanov2017d, Ivanov2019}). As a result, taking into account the
constraints on the scalar interaction \cite{Hardy2015} and
\cite{Severijns2019} the contributions of interactions BSM to the
neutron $\beta^-$--decay can be induced only by a tensor nucleon
current \cite{Pattie2013, Ivanov2018c}. As we show below the
contribution of the one--pion--pole (OPP) exchange to the correlation
coefficients of the neutron $\beta^-$--decays for a polarized
neutron, a polarized electron and an unpolarized proton is of order
$10^{-5}$. This is commensurable with the contribution of the isospin
breaking correction to the vector coupling constant of the neutron
$\beta^-$--decay calculated by Kaiser \cite{Kaiser2001} within the
heavy baryon chiral perturbation theory (HB$\chi$PT). However, unlike
Kaiser's correction the contribution of the OPP exchange can be
screened by the contribution of the pseudoscalar interaction BSM.
 
This paper is addressed to the analysis of contributions of the OPP
exchange, caused by strong low--energy interactions, and the
pseudoscalar interaction BSM introduced in
\cite{Lee1956}--\cite{Severijns2006} to the neutron lifetime and
correlation coefficients of the neutron $\beta^-$--decays for a
polarized neutron, a polarized electron and unpolarized proton. The
analysis of contributions of pseudoscalar interactions to the
electron--energy and angular distribution of the neutron
$\beta^-$--decay for a polarized neutron and unpolarized electron and
proton has a long history \cite{Harrington1960}--\cite{Hayen2018} (see
also \cite{Wilkinson1982, Severijns2019}). For example the Fierz--like
interference term \cite{Fierz1937}, induced by pseudoscalar
interactions, can be recognized in the electron--energy and angular
distributions calculated in \cite{Harrington1960}--\cite{Hayen2018}
(see also \cite{Wilkinson1982, Severijns2019}). The contributions of
the pseudoscalar interactions to the correlation coefficients of the
electron--energy and angular distribution of the neutron
$\beta^-$--decay for a polarized neutron and unpolarized electron and
proton can be, in principle, extracted from the electron--energy and
angular distributions obtained by Harrington \cite{Harrington1960}
(see Eqs.(9) -- (13) of Ref.\cite{Harrington1960}) and Holstein
\cite{Holstein1974} (see Appendix B of Ref.\cite{Holstein1974}) (see
also section \ref{sec:schluss} of this paper). In our work in addition
to the results obtained in \cite{Harrington1960}--\cite{Hayen2018}
(see also \cite{Wilkinson1982, Severijns2019}) we calculate the
contributions of pseudoscalar interactions to the correlation
coefficients of the electron--energy and angular distribution of the
neutron $\beta^-$--decays, caused by correlations with the electron
spin. The analyze of contributions of pseudoscalar interactions to the
correlation coefficients of the electron--energy and angular
distribution of the neutron $\beta^-$--decays for a polarized neutron,
a polarized electron and unpolarized proton, carried out in this
paper, completes the investigations of contributions of interactions
BSM to the electron--energy and angular distributions, which we have
performed in \cite{Ivanov2017b, Ivanov2017d, Ivanov2019}, where we
have calculated i) the complete set of corrections of order $10^{-3}$,
caused by radiative corrections of order $O(\alpha/\pi)$ and the weak
magnetism and proton recoil corrections of order $O(E_e/M)$, and ii)
contributions of vector, axial--vector, scalar and tensor interactions
BSM introduced in \cite{Lee1956}--\cite{Severijns2006}.

The paper is organized as follows. In section \ref{sec:amplitude} we
write down the amplitude of the neutron $\beta^-$--decay by taking
into account the contributions of the OPP exchange and the
pseudoscalar interaction BSM only. We analyze the contributions of
energy independent corrections to the pseudoscalar form factor of the
nucleon defined by the Adler-Dothan-Wolfenstein (ADM) term
\cite{Adler1966, Wolfenstein1970} and chiral corrections calculated
within the HB$\chi$PT \cite{Bernard1995, Bernard1996, Kaiser2003}. We
show that the ADM--term and chiral corrections, calculated in the
two--loop approximation within the HB$\chi$PT by Kaiser
\cite{Kaiser2003}, are able in principle to induce sufficiently small
real contributions to phenomenological coupling constants of the
pseudoscalar interaction BSM of a neutron--proton pseudoscalar density
coupled to a left--handed leptonic current. In section
\ref{sec:spectrum} we discuss the contributions to the correlation
coefficients of the electron--energy and angular distribution of the
neutron $\beta^-$--decays caused by the OPP exchange and the
pseudoscalar interaction BSM. The distribution is calculated for a
polarized neutron, a polarized electron and an unpolarized proton.
Using the results, obtained in \cite{Bhattacharya2012, Severijns2019,
  Gonzalez-Alonso2014,Gonzalez-Alonso2016} we estimate the
phenomenological coupling constants of the pseudoscalar interactions
BSM. We adduce the results in Table I. In section \ref{sec:schluss} we
discuss the obtained results, which can be used for experimental
analyses of the neutron $\beta^-$--decays with experimental accuracies
of about a few parts of $10^{-5}$ \cite{Abele2016}. Since the complete
set of contributions of order $10^{-3}$, including the radiative
corrections of order $O(\alpha/\pi)$ and corrections of order
$O(E_0/M)$, caused by the weak magnetism and proton recoil, are
calculated at the neglect of contributions of order $O(\alpha E_0/\pi
M) \sim 10^{-6}$ and $O(E^2_0/M^2) \sim 10^{-6}$ \cite{Ivanov2013,
  Ivanov2017b, Ivanov2017d, Ivanov2019}, the results obtained in this
paper should be tangible and important for a correct analysis of
experimental data on searches of contributions of interactions BSM
with an accuracy of a few parts of $10^{-5}$. We give also a
comparative analysis of the results obtained in this work with those
in \cite{Wilkinson1982, Harrington1960}--\cite{Hayen2018}. This allows
us to argue that the corrections, caused by pseudoscalar interactions,
calculated for the correlation coefficients of the neutron
$\beta^-$--decays, induced by correlations of the electron spin with
the neutron spin and 3-momenta of decay fermions with standard
correlation structures introduced by Jackson {\it et al.}
\cite{Jackson1957}, are fully new. Moreover all terms in
Eq.(\ref{eq:A.6}) with correlation structures beyond the standard ones
by Jackson {\it et al.}  \cite{Jackson1957} and proportional to the
effective coupling constants $C'_{ps}$ and $C''_{ps}$ were never
calculated in literature. In the Appendix we give a detailed
calculation of the contributions of pseudoscalar interactions caused
by the OPP exchange and BSM to the correlation coefficients of the
neutron $\beta^-$--decays for a polarized neutron, a polarized
electron and an unpolarized proton, completing the analysis of
contributions of interactions BSM to the correlation coefficients of
the neutron $\beta^-$--decays carried out in \cite{Ivanov2017b,
  Ivanov2017d, Ivanov2019}.

\section{Amplitude of the neutron $\beta^-$--decay with contributions
  of OPP exchange and pseudoscalar interaction BSM}
\label{sec:amplitude}

Since the expected order of contributions of pseudoscalar interactions
of about $10^{-5}$, we take them into account in the linear
approximation additively to the corrections of order $10^{-4} -
10^{-3}$ calculated in \cite{Bilenky1959}--\cite{Severijns2019}. In
such an approximation and following \cite{Ivanov2013, Ivanov2017d,
  Ivanov2019} the amplitude of the neutron $\beta^-$--decay we take in
the form
\begin{eqnarray}\label{eq:1}
  M(n \to p e^- \bar{\nu}_e) &=& -
  \frac{G_F}{\sqrt{2}}\,V_{ud}\,\Big\{\langle p(\vec{k}_p,
  \sigma_p)|J^{(+)}_{\mu}(0)|n(\vec{k}_n,
  \sigma_n)\rangle\,\big[\bar{u}_e(\vec{k}_e, \sigma_e)\gamma^{\mu}(1
    - \gamma^5)v_{\bar{\nu}}(\vec{k}_{\bar{\nu}}, +
    \frac{1}{2})\big]\nonumber\\ &&\hspace{0.50in}~ +
  \bar{u}_p(\vec{k}_p, \sigma_p)\gamma^5 u_n(\vec{k}_n,
  \sigma_n)\,\big[\bar{u}_e(\vec{k}_e, \sigma_e) (C_p + \bar{C}_P
    \gamma^5)v_{\bar{\nu}}(\vec{k}_{\bar{\nu}}, +
    \frac{1}{2})\big]\Big\},
\end{eqnarray}
where $G_F$ and $V_{ud}$ are the Fermi couping constant and the
Cabibbo--Kobayashi--Maskawa (CKM) matrix element \cite{PDG2018}. Then,
$\langle p(\vec{k}_p, \sigma_p)|J^{(+)}_{\mu}(0)|n(\vec{k}_n,
\sigma_n)\rangle$ is the matrix element of the charged hadronic
current $J^{(+)}_{\mu}(0) = V^{(+)}_{\mu}(0) - A^{(+)}_{\mu}(0)$,
where $V^{(+)}_{\mu}(0)$ and $A^{(+)}_{\mu}(0)$ are the charged vector
and axial--vector hadronic currents \cite{Feynman1958, Nambu1960,
  Marshak1969}. The fermions in the initial and final states are
described by Dirac bispinor wave functions $u_n$, $u_p$, $u_e$ and
$v_{\bar{\nu}}$ of free fermions \cite{Ivanov2013, Ivanov2014}. In the
second term of Eq.(\ref{eq:1}) we take into account the contribution
of the pseudoscalar interaction BSM
\cite{Lee1956}--\cite{Severijns2006} with two complex phenomenological
coupling constants $C_P$ and $\bar{C}_P$ in the notation of
\cite{Ivanov2013, Ivanov2017d, Ivanov2019}.

For the analysis of contributions of pseudoscalar interactions to the
neutron $\beta^-$--decays for a polarized neutron, a polarized
electron and an unpolarized proton we define the matrix element
$\langle p(\vec{k}_p, \sigma_p)|J^{(+)}_{\mu}(0)|n(\vec{k}_n,
\sigma_n)\rangle$ as follows
\begin{eqnarray}\label{eq:2}
  \langle p(\vec{k}_p, \sigma_p)|J^{(+)}_{\mu}(0)|n(\vec{k}_n,
  \sigma_n)\rangle = \bar{u}_p(\vec{k}_p,
  \sigma_p)\Big(\gamma_{\mu}(1 + \lambda \gamma^5) + \frac{2 M \lambda
    \, q_{\mu}}{m^2_{\pi} - q^2 - i0}\gamma^5\Big)u_n(\vec{k}_n,
  \sigma_n),
\end{eqnarray}
where $\lambda$ is the axial coupling constant with recent
experimental value $\lambda = - 1.27641(45)_{\rm stat.}(33)_{\rm
  syst.}$ \cite{Abele2018}.  The first term in Eq.(\ref{eq:1}) is
written in agreement with the standard $V - A$ effective theory of
weak interactions \cite{Feynman1958,Nambu1960, Marshak1969} (see also
\cite{Shekhter1959a, Shekhter1959b}). The term
proportional to $q_{\mu} \gamma^5$ defines the contribution of the OPP
exchange, caused by strong low--energy interactions (see also
\cite{Nambu1960}) with the $\pi^-$--meson mass $m_{\pi} =
139.57061(24)\,{\rm MeV}$ \cite{PDG2018} and $q = k_p - k_n = - k_e -
k_{\bar{\nu}}$ is a 4--momentum transfer. The OPP contribution is
required by conservation of the charged hadronic axial--vector current
in the chiral limit $m_{\pi} \to 0$ \cite{Nambu1960}.

In the more general form the matrix element of the hadronic
axial--vector current can be taken in the form accepted in the
HB$\chi$PT \cite{Bernard1995, Bernard1996, Kaiser2003}. This gives
\begin{eqnarray}\label{eq:3}
  \langle p(\vec{k}_p, \sigma_p)|A^{(+)}_{\mu}(0)|n(\vec{k}_n,
  \sigma_n)\rangle = \bar{u}_p(\vec{k}_p,
  \sigma_p)\Big(\gamma_{\mu}\,G_A(q^2) +
  \frac{q_{\mu}}{2M}\,G_P(q^2)\Big)\,\gamma^5 u_n(\vec{k}_n, \sigma_n),
\end{eqnarray}
where $G_A(q^2)$ and $G_P(q^2)$ are the axial--vector form factor and
the induced pseudoscalar form factor, respectively, at $0 \le q^2 \le
\Delta^2$ for the neutron $\beta^-$--decay with $\Delta = m_n -
m_p$. The invariant 4--momentum transfer squared $q^2$ vanishes,
i.e. $q^2 = 0$, at the kinetic energy of the proton $T_p = E_p - m_p =
\Delta^2/2 m_n$. In the chiral limit $m_{\pi} \to 0$ because of
conservation of the charged hadronic axial--vector current
\cite{Nambu1960} the form factors $G_A(q^2)$ and $G_P(q^2)$ are
related by $G_P(q^2) = - (4M^2/q^2) G_A(q^2)$. In turn, for a finite
pion mass the pseudoscalar form factor $G_P(q^2)$ has been calculated
in the two--loop approximation within HB$\chi$PT by Kaiser
\cite{Kaiser2003}. A precision analysis of the induced pseudoscalar
form factor in the proton weak interactions has been also carried out
by Gorringe and Fearing \cite{Gorringe2004}.

\subsection{Pseudoscalar interaction BSM as induced by corrections
  to the pseudoscalar form factor, caused by strong low--energy
  interactions}

 According to \cite{Bernard1995}, the axial--vector form factor
$G_A(q^2)$ can be rather good parameterized by a dipole form (see also
\cite{Liesenfeld1999})
\begin{eqnarray}\label{eq:4}
  G_A(q^2) = \frac{g_A}{\big(1 + q^2/M^2_A\big)^2} = g_A\Big(1 -
  \frac{1}{6}\,\langle r^2_A\rangle q^2 + \ldots\Big),
\end{eqnarray}
where $g_A = - \lambda$ is the axial--coupling constant, and $M_A$ is
the cut--off mass related to the mean square axial radius of the
nucleon $\langle r^2_A\rangle$ as $\langle r^2_A\rangle = 12/M^2_A =
0.403(29)\,{\rm fm^2}$ with $M_A = 1.077(39)\,{\rm GeV}$ extracted
from charged pion electroproduction experiments
\cite{Liesenfeld1999}. In turn, the cut--off mass $M_A =
1.026(17)\,{\rm GeV}$ extracted from (quasi)elastic neutrino and
antineutrino scattering experiments \cite{Liesenfeld1999} gives
$\langle r^2_A\rangle = 12/M^2_A = 0.440(16)\,{\rm fm^2}$. In the
approximation Eq.(\ref{eq:4}) the pseudoscalar form factor $G_P(q^2)$
acquires the following form \cite{Bernard1995} (see also
\cite{Gorringe2004})
\begin{eqnarray}\label{eq:5}
  \frac{1}{2M\,}G_P(q^2) = \frac{2M g_A}{m^2_{\pi} - q^2 - i0} -
  \frac{1}{3}\,g_A M \langle r^2_A\rangle,
\end{eqnarray}
where the correction to the OPP exchange is the
Adler--Dothan--Wolfenstein (ADW) term \cite{Adler1966,
  Wolfenstein1970}. The ADW--term induces the BSM--like pseudoscalar
interaction with the coupling constants
\begin{eqnarray}\label{eq:6}
 C^{(\rm ADW)}_P = - \bar{C}^{(\rm ADW)}_P = - \frac{1}{3}\,\lambda
 \langle r^2_A\rangle\,m_e M = 2.1\times 10^{-3}.
\end{eqnarray}
According to Eq.(\ref{eq:11}), this gives the contribution to the
correlation coefficients of the neutron $\beta^-$--decays equal to
${\rm Re}\,C^{(\rm BSM)}_{ps} = C^{(\rm ADW)}_{ps} = - 4.9 \times
10^{-7}$. Using the results, obtained by Kaiser \cite{Kaiser2003} (see
Eq.(7) of Ref.\cite{Kaiser2003}) in the two--loop approximation in the
HB$\chi$PT, the induced BSM--like pseudoscalar coupling constants are
equal to
\begin{eqnarray}\label{eq:7}
 C^{(\rm K)}_P = - \bar{C}^{(\rm K)}_P = \frac{m_e m^2_{\pi}M}{32
   \pi^4 f^4_{\pi}}\,\zeta_0 = 4.1\times 10^{-5}\,\zeta_0,
\end{eqnarray}
where $f_{\pi} = 92.4\,{\rm MeV}$ is the charged pion leptonic (or
PCAC) constant \cite{Bernard1995, Kaiser2003}.  Since $|\zeta_0|\sim
1$ \cite{Kaiser2003}, we get $|C_P| = |\bar{C}_P|\sim 4.1\times
10^{-5}$. The contribution of $C^{(\rm K)}_P = - \bar{C}^{(\rm K)}_P$
to the coupling constant ${\rm Re}\,C^{(\rm BSM)}_{ps}$ (see
Eq.(\ref{eq:11})) is of order $|{\rm Re}\,C^{(\rm BSM)}_{ps}| \sim
9.6\times 10^{-9}$. This means that the SM strong low--energy
interactions are able to induce the BSM--like pseudoscalar interaction
with real coupling constants, the contributions of which are much
smaller than the current experimental sensitivity of the neutron
$\beta^-$--decays \cite{Abele2016}.  Below we consider a more general
pseudoscalar interaction BSM with complex phenomenological coupling
constants $C_P$ and $\bar{C}_P$ such as $C_P \neq - \bar{C}_P$.

\subsection{Non--relativistic approximation for the amplitude of the
  neutron $\beta^-$--decay Eq.(\ref{eq:1})}

In the non--relativistic approximation for the neutron and proton the
amplitude of the neutron $\beta^-$--decay in Eq.(\ref{eq:1}) takes the
form
\begin{eqnarray}\label{eq:8}
 M(n \to p e^- \bar{\nu}_e) &=& - \frac{G_F}{\sqrt{2}}\,V_{ud}2 M
 \Big\{[\varphi^{\dagger}_p\varphi_n][\bar{u}_e\gamma^0(1 -
   \gamma^5)v_{\bar{\nu}}] -
 \lambda[\varphi^{\dagger}_p\vec{\sigma}\,\varphi_n]\cdot [\bar{u}_e
   \vec{\gamma}\,(1 - \gamma^5)v_{\bar{\nu}}]\nonumber\\ && +
 \lambda\,\frac{m_e}{m^2_{\pi}}[\varphi^{\dagger}_p(\vec{\sigma}\cdot
   \vec{k}_p)\varphi_n][\bar{u}_e(1 - \gamma^5)v_{\bar{\nu}}] -
 \frac{1}{2 M}[\varphi^{\dagger}_p(\vec{\sigma}\cdot
   \vec{k}_p)\varphi_n][\bar{u}_e(C_p + \bar{C}_P
   \gamma^5)v_{\bar{\nu}}]\Big\},
\end{eqnarray}
where $\varphi_j$ for $j = p,n$ are the Pauli spinorial wave functions
of non--relativistic neutron and proton, and $\vec{k}_p = - \vec{k}_e
- \vec{k}_{\bar{\nu}}$ is a 3--momentum of the proton.

\section{Electron--energy and angular distribution of the neutron
  $\beta^-$--decay for polarized neutron, polarized electron, and
  unpolarized proton}
\label{sec:spectrum}

The electron--energy and angular distribution of the neutron
$\beta^-$--decays for a polarized neutron, a polarized electron and an
unpolarized proton has been written by Jackson {\it et al.}
\cite{Jackson1957}. It reads
\begin{eqnarray}\label{eq:9}
\hspace{-0.15in}&&\frac{d^5 \lambda_n(E_e, \vec{k}_e,
  \vec{k}_{\bar{\nu}}, \vec{\xi}_n, \vec{\xi}_e)}{dE_e d\Omega_e
  d\Omega_{\bar{\nu}}} = (1 + 3
\lambda^2)\,\frac{G^2_F|V_{ud}|^2}{32\pi^5}\,(E_0 - E_e)^2
\,\sqrt{E^2_e - m^2_e}\, E_e\,F(E_e, Z = 1)\,\zeta(E_e)\,\Big\{1 +
b\,\frac{m_e}{E_e}\nonumber\\
\hspace{-0.15in}&& + a(E_e)\,\frac{\vec{k}_e\cdot
  \vec{k}_{\bar{\nu}}}{E_e E_{\bar{\nu}}} +
A(E_e)\,\frac{\vec{\xi}_n\cdot \vec{k}_e}{E_e} + B(E_e)\,
\frac{\vec{\xi}_n\cdot \vec{k}_{\bar{\nu}}}{E_{\bar{\nu}}} +
K_n(E_e)\,\frac{(\vec{\xi}_n\cdot \vec{k}_e)(\vec{k}_e\cdot
  \vec{k}_{\bar{\nu}})}{E^2_e E_{\bar{\nu}}}+
Q_n(E_e)\,\frac{(\vec{\xi}_n\cdot \vec{k}_{\bar{\nu}})(\vec{k}_e\cdot
  \vec{k}_{\bar{\nu}})}{ E_e E^2_{\bar{\nu}}}\nonumber\\
\hspace{-0.15in}&& + D(E_e)\,\frac{\vec{\xi}_n\cdot (\vec{k}_e\times
  \vec{k}_{\bar{\nu}})}{E_e E_{\bar{\nu}}} + G(E_e)\,\frac{\vec{\xi}_e
  \cdot \vec{k}_e}{E_e} + H(E_e)\,\frac{\vec{\xi}_e \cdot
  \vec{k}_{\bar{\nu}}}{E_{\bar{\nu}}} + N(E_e)\,\vec{\xi}_n\cdot
\vec{\xi}_e + Q_e(E_e)\,\frac{(\vec{\xi}_n\cdot \vec{k}_e)(
  \vec{k}_e\cdot \vec{\xi}_e)}{(E_e + m_e) E_e}\nonumber\\
\hspace{-0.15in}&& + K_e(E_e)\,\frac{(\vec{\xi}_e\cdot \vec{k}_e)(
  \vec{k}_e\cdot \vec{k}_{\bar{\nu}})}{(E_e + m_e)E_e E_{\bar{\nu}}} +
R(E_e)\,\frac{\vec{\xi}_n\cdot(\vec{k}_e \times \vec{\xi}_e)}{E_e} +
L(E_e)\,\frac{\vec{\xi}_e\cdot(\vec{k}_e \times
  \vec{k}_{\bar{\nu}})}{E_eE_{\bar{\nu}}} - 3\,\frac{E_e}{M}\,\frac{1
  - \lambda^2}{1 + 3 \lambda^2}\,\Big(\frac{(\vec{k}_e\cdot
  \vec{k}\,)^2}{E^2_e E^2} - \frac{1}{3}\,\frac{k^2_e}{E^2_e}\Big)
\nonumber\\
\hspace{-0.3in}&& + 3\,\frac{1 - \lambda^2}{1 + 3 \lambda^2}\,
\frac{m_e}{M}\,\Big(\frac{(\vec{\xi}_e\cdot
  \vec{k}_{\nu})(\vec{k}_e\cdot \vec{k}_{\nu})}{E_e E^2_{\nu}} -
\frac{1}{3}\,\frac{\vec{\xi}_e\cdot \vec{k}_e}{E_e}\,\Big) +
3\,\frac{1 - \lambda^2}{1 + 3 \lambda^2}\,
\frac{1}{M}\,\Big(\frac{(\vec{\xi}_e\cdot \vec{k}_e)(\vec{k}_e\cdot
  \vec{k}_{\nu})^2}{(E_e + m_e)E_e E^2_{\nu}} - \frac{1}{3}\,(E_e -
m_e)\,\frac{\vec{\xi}_e\cdot \vec{k}_e}{E_e}\,\Big) \Big\},
\end{eqnarray}
where we have followed the notation \cite{Ivanov2013, Ivanov2017b,
  Ivanov2017d, Ivanov2019}. The last three terms in Eq.(\ref{eq:9})
are caused by the contributions of the proton recoil calculated to
order $O(E_e/M)$ \cite{Gudkov2006, Ivanov2013, Ivanov2017b,
  Ivanov2017d, Ivanov2019}. Then, $\vec{\xi}_n$ and $\vec{\xi}_e$ are
unit polarization vectors of the neutron and electron, respectively,
$d\Omega_e$ and $d\Omega_{\bar{\nu}}$ are infinitesimal solid angels
in the directions of electron $\vec{k}_e$ and antineutrino
$\vec{k}_{\bar{\nu}}$ 3--momenta, respectively, $E_0 = (m^2_n - m^2_p
+ m^2_e)/2m_n = 1.2926\,{\rm MeV}$ is the end--point energy of the
electron spectrum, $F(E_e, Z = 1)$ is the relativistic Fermi function
equal to \cite{Blatt1952}--\cite{Konopinski1966} (see also
\cite{Wilkinson1982, Ivanov2013, Ivanov2017b, Ivanov2017d,
  Ivanov2019})
\begin{eqnarray}\label{eq:10}
\hspace{-0.3in}F(E_e, Z = 1 ) = \Big(1 +
\frac{1}{2}\gamma\Big)\,\frac{4(2 r_pm_e\beta)^{2\gamma}}{\Gamma^2(3 +
  2\gamma)}\,\frac{\displaystyle e^{\,\pi
 \alpha/\beta}}{(1 - \beta^2)^{\gamma}}\,\Big|\Gamma\Big(1 + \gamma +
 i\,\frac{\alpha }{\beta}\Big)\Big|^2,
\end{eqnarray}
where $\beta = k_e/E_e = \sqrt{E^2_e - m^2_e}/E_e$ is the electron
velocity, $\gamma = \sqrt{1 - \alpha^2} - 1$, $r_p$ is the electric
radius of the proton.  In the numerical calculations we use $r_p =
0.841\,{\rm fm}$ \cite{Pohl2010}. The function $\zeta(E_e)$ contains
the contributions of radiative corrections of order $O(\alpha/\pi)$
and corrections from the weak magnetism and proton recoil of order
$O(E_e/M)$, taken in the form used in \cite{Gudkov2006, Ivanov2013,
  Ivanov2017b, Ivanov2017d, Ivanov2019}. Then, $b$ is the Fierz
interference term defined by the contributions of interactions beyond
the SM \cite{Fierz1937}. The analytical expressions for the
correlation coefficients $a(E_e)$, $A(E_e)$ and so on, calculated
within the SM with the account for radiative corrections of order
$O(\alpha/\pi)$ and corrections caused by the weak magnetism and
proton recoil of order $O(E_e/M)$ together with the contributions of
Wilkinson's corrections \cite{Wilkinson1982}, are given in
\cite{Ivanov2013, Ivanov2017b, Ivanov2017d, Ivanov2019}.

\subsection{Corrections to the correlation coefficients of the
  electron--energy and angular distribution of the neutron
  $\beta^-$--decays caused by pseudoscalar interactions}

In the Appendix we calculate the contributions of the OPP exchange and
the pseudoscalar interaction BSM to the correlation coefficients of
the electron--energy and angular distribution of the neutron
$\beta^-$--decays for a polarized neutron, a polarized electron and
an unpolarized proton. The corrections to the correlation coefficients
and the correction to the electron--energy and angular distribution
are given in the Appendix in Eqs.(\ref{eq:A.5}) and (\ref{eq:A.6}),
respectively.  The strength of these corrections (see
Eq.(\ref{eq:A.5})) is defined by the effective coupling constants
$C'_{ps}$ and $C''_{ps}$, which are the real and imaginary parts of
the effective coupling constant $C_{ps}$ given by
\begin{eqnarray}\label{eq:11}
 C_{ps} &=& C^{(\rm OPP)}_{ps} + C^{(\rm BSM)}_{ps} = C'_{ps} +
 i\,C''_{ps},\nonumber\\ C^{(\rm OPP)}_{ps} &=& \frac{2\lambda}{1 +
   3\lambda^2}\,\frac{m_e }{m^2_{\pi}}\,E_0 = - 1.47\times
 10^{-5},\nonumber\\ C^{(\rm BSM)}_{ps} &=& - \frac{1}{1 +
   3\lambda^2}\, \frac{E_0}{2M}\,(C_P - \bar{C}_P) = - 1.17\times
 10^{-4}\,(C_P - \bar{C}_P),\nonumber\\ C'_{ps} &=& {\rm Re}\,C_{ps} =
 C^{(\rm OPP)}_{ps} + {\rm Re}\, C^{(\rm
   BSM)}_{ps},\nonumber\\ C''_{ps} &=& {\rm Im}\,C_{ps} = {\rm
   Im}\,C^{(\rm BSM)}_{ps},
\end{eqnarray}
where $C^{(\rm OPP)}_{ps}$ and $C^{(\rm BSM)}_{ps}$ are the effective
coupling constants caused by the OPP exchange and the pseudoscalar
interaction BSM, respectively. The numerical values are calculated for
$\lambda = - 1.27641$ \cite{Abele2018}, $m_e = 0.5110\,{\rm MeV}$,
$m_{\pi} = 139.5706\,{\rm MeV}$ \cite{PDG2018}, $E_0 = (m^2_n - m^2_p
+ m^2_e)/2m_n = 1.2926\,{\rm MeV}$ and $M = (m_n + m_p)/2 =
938.9188\,{\rm MeV}$ \cite{PDG2018}, respectively. According to our
analysis (see Eqs.(\ref{eq:6}) and (\ref{eq:7})), a real part of the
phenomenological coupling constant $C^{(\rm BSM)}_{ps}$ can be partly
induced by the SM strong low--energy interactions through the
ADM--term (see Eq.(\ref{eq:6})) and Kaiser's two--loop corrections,
calculated within the HB$\chi$PT (see Eq.(\ref{eq:7})).

The corrections, caused by pseudoscalar interactions (see
Eq.(\ref{eq:A.5}) and Eq.(\ref{eq:A.6})), to the electron--energy and
angular distribution of the neutron $\beta^-$--decays for a polarized
neutron, a polarized electron and an unpolarized proton, taken
together with the electron--energy and angular distributions
calculated in \cite{Gudkov2006, Ivanov2013, Ivanov2017b, Ivanov2017d,
  Ivanov2019} can be used as a theoretical background for experimental
searches of contributions of interactions BSM of order $10^{-4}$ or
even smaller \cite{Abele2016}.

\subsection{Estimates of the real and imaginary parts of the
  phenomenological coupling constant $C_P - \bar{C}_P$}

According to \cite{Cirigliano2013}, the phenomenological coupling
constant $C_P - \bar{C}_P$ can be defined as follows
\begin{eqnarray}\label{eq:12}
C_P - \bar{C}_P = 2\, g_P\,\epsilon_P,
\end{eqnarray}
where $\epsilon_P$ is a complex effective coupling constant of the
four--fermion local weak interaction of the pseudoscalar quark current
$\bar{u}\gamma^5 d$, where $u$ and $d$ are the {\it up} and {\it
  down} quarks, with the left--handed leptonic current $ \bar{\ell}(1
- \gamma^5)\nu_{\ell}$ \cite{Cirigliano2010} -- \cite{Cirigliano2013a}
(see also \cite{Gonzalez-Alonso2014, Severijns2019}). Then, $g_P$ is
the matrix element $\langle p|\bar{u}\gamma^5 d|n\rangle =
g_P\bar{u}_p\gamma^5 u_n$ caused by strong low--energy interactions,
where $\bar{u}_p$ and $u_n$ are the Dirac wave functions of a free
proton and neutron, respectively. According to Gonz\'alez-Alonso and
Camalich \cite{Gonzalez-Alonso2014}, one gets $g_P = 349(9)$ (see
Eq.(13) of Ref.\cite{Gonzalez-Alonso2014}).

\begin{table}[h]
\begin{tabular}{|c|c|}
  \hline $0 \lesssim {\rm Re}\,(C_p - \bar{C}_P) \lesssim 0.3 $ & $ -
  3.5 \times 10^{-5} \lesssim {\rm Re}\,C^{(\rm BSM)}_{ps} \lesssim 0$
  \\\hline ${\rm Im}\,(C_P - \bar{C}_P) < 0.2$ & $ {\rm Im}\,C^{(\rm
    BSM)}_{ps} < - 2.3 \times 10^{-5}$ \\ \hline
\end{tabular} 
\caption{Estimates of the phenomenological coupling constant $C_P -
  \bar{C}_p = 2g_P\,\epsilon_P$ for $g_P = 349(9)$
  \cite{Gonzalez-Alonso2014} and the constraints on the parameter
  $\epsilon_P$ \cite{Bhattacharya2012, Severijns2019,
    Gonzalez-Alonso2014, Gonzalez-Alonso2016}.}
\end{table}

Following \cite{Gonzalez-Alonso2014} and using the constraint
$|\epsilon_P| < 5.8\times 10^{-3}$, obtained at $90\,\%$ C.L. from the
experimental data on the search for an excess of events with a charged
lepton (an electron or muon) and a neutrino in the final state of the
pp collision with the centre-of-mass energy of $\sqrt{s} = 8\,{\rm
  TeV}$ with an integrated luminosity of $20\,{\rm fb^{-1}}$ at LHC
\cite{CMS2013}, we get $|{\rm Re}(C_P - \bar{C}_P)| < 4.1$. In this
case the pseudoscalar interaction BSM can dominate in the effective
coupling constant $C'_{ps}$ in comparison to the OPP exchange, which
is of order $|C^{(\rm OPP)}_{ps}| \sim 10^{-5}$.

In turn, the analysis of the leptonic decays of charged pions, carried
out in \cite{Severijns2019} (see Eq.(113) and a discussion on p.51 of
Ref.\cite{Severijns2019}), taken together with the results, obtained
in \cite{Gonzalez-Alonso2016}, gives one ${\rm Re}\,\epsilon_P =
(0.4\pm 1.3)\times 10^{-4}$ and, correspondingly, ${\rm Re}(C_P -
\bar{C}_P) = 0.03 \pm 0.09$. Such an analysis implies that the
phenomenological coupling constants ${\rm Re}(C_P - \bar{C}_P)$ and
$C^{(\rm BSM)}_{ps}$ are commensurable with zero. This leads to a
dominate role of the OPP exchange in the effective coupling constant
$C'_{ps}$ equal to $C'_{ps} = - 1.47 \times 10^{-5}$.

Then, following the assumption $\epsilon_P = 2 m_e(m_u +
m_d)/m^2_{\pi} \sim 4\times 10^{-4}$ \cite{Severijns2019},which is
also related to the analysis of the leptonic decays of charged pions
(see a discussion below Eq.(112) of Ref.\cite{Severijns2019}), we get
${\rm Re}(C_P - \bar{C}_P) \sim 0.3$ and ${\rm Re}\,C^{(\rm BSM)}_{ps}
\sim - 3.5 \times 10^{-5}$. As a result, according to the assumption
$\epsilon_P = 2 m_e(m_u + m_d)/m^2_{\pi} \sim 4\times 10^{-4}$, the
contribution of the pseudoscalar interaction BSM to the effective
coupling constant $C'_{ps}$ should be of order $10^{-5}$, that makes
it commensurable with the contribution of the OPP exchange.

Since the constraint $|\epsilon_P| < 5.8 \times 10^{-3}$
\cite{Gonzalez-Alonso2014} disagrees with the constraints following
from the analysis of the leptonic decays of charged pions
\cite{Severijns2019, Gonzalez-Alonso2016}, one may conclude that the
phenomenological coupling constant ${\rm Re}(C_P - \bar{C}_P)$ should
be constrained by $ 0 \lesssim {\rm Re}(C_P - \bar{C}_P) \lesssim
0.3$.  This leads to the effective coupling constant ${\rm
  Re}\,C^{(\rm BSM)}_{ps}$ restricted by $ - 3.5 \times 10^{-5}
\lesssim {\rm Re}\,C^{(\rm BSM)}_{ps} \lesssim 0$. This shifts the
contributions of the pseudoscalar interaction BSM to the region of
values $|{\rm Re}\,C^{(\rm BSM)}_{ps}| \sim 10^{-5}$ or even smaller.

The imaginary part ${\rm Im}(C_P - \bar{C}_P) = 2 g_P \,{\rm
  Im}\,\epsilon_P$ we estimate using the upper bound ${\rm
  Im}\,\epsilon_P < 2.8 \times 10^{-4}$, obtained at $90\,\%$ C.L. in
\cite{Bhattacharya2012} (see also Eq.(114) of
Ref.\cite{Severijns2019}). We get ${\rm Im}(C_P - \bar{C}_P) <
0.3$. The effective coupling constant $C''_{ps} = {\rm Im}\,C^{(\rm
  BSM)}_{ps}$ is restricted by $C''_{ps} = {\rm Im}\,C^{(\rm
  BSM)}_{ps} < - 2.3 \times 10^{-5}$. Since the contribution of the
OPP exchange is real, the effective coupling constant $C''_{ps}$,
constrained by $C''_{ps} < - 2.3\times 10^{-5}$, is fully defined by
the pseudoscalar interaction BSM.

In Table I we adduce the constraints on the real and imaginary parts
of the phenomenological coupling constant $C_P - \bar{C}_P$ and on the
effective coupling constant $C^{(\rm BSM)}_{ps}$, which may follow
from the results obtained in \cite{Severijns2019,
  Gonzalez-Alonso2014,Gonzalez-Alonso2016}.

\section{Discussion}
\label{sec:schluss}

The corrections of order $10^{-5}$, calculated within the SM, are
needed as a SM theoretical background for experimental searches of
interactions beyond the SM in terms of asymmetries and correlation
coefficients of the neutron $\beta^-$--decays \cite{Ivanov2017b,
  Ivanov2017d, Ivanov2019}. An experimental accuracy of about a few
parts of $10^{-5}$ or even better, which is required for experimental
analyses of interactions BSM of order $10^{-4}$, can be reachable at
present time \cite{Abele2016}. In this paper we have continued the
analysis of corrections of order $10^{-5}$ to the correlation
coefficients of the neutron $\beta^-$--decays, which we have begun in
\cite{Ivanov2017b, Ivanov2017d, Ivanov2019, Ivanov2019a}.
In this
paper we have taken into account the contributions of strong
low--energy interactions in terms of the OPP exchange and the
contributions of the pseudoscalar interaction BSM
\cite{Lee1956}--\cite{Severijns2006}, and calculated corrections to
the correlation coefficients of the electron--energy and angular
distribution of the neutron $\beta^-$--decay for a polarized neutron,
a polarized electron and an unpolarized proton.

In addition to the results, concerning the corrections caused by
pseudoscalar interactions to the electron--energy and angular
distributions of the neutron $\beta^-$--decay for a polarized neutron
and unpolarized electron and proton, obtained in
\cite{Harrington1960}--\cite{Hayen2018} and especially by Harrington
\cite{Harrington1960} and Holstein \cite{Holstein1974}, we have
calculated corrections to the correlation coefficients, caused by
correlations with the electron spin, i.e. for a polarized neutron and
a polarized electron with an unpolarized proton.

We have shown that the energy independent contributions to the
pseudoscalar form factor \cite{Bernard1995, Bernard1996, Kaiser2003,
  Adler1966, Wolfenstein1970}, related to the Adler-Dothan-Wolfenstein
(ADM) term Eq.(\ref{eq:6}) and to the chiral corrections
Eq.(\ref{eq:7}), calculated by Kaiser \cite{Kaiser2003} in a two--loop
approximation within the HB$\chi$PT, are able in principle to be
responsible for sufficiently small real parts of the phenomenological
coupling constants $C_P$ and $\bar{C}_P$ and at the level of $10^{-6}
- 10^{-8}$ of the effective coupling constant $C^{(\rm BSM)}_{ps}$. In
turn, the isospin breaking corrections of order $10^{-5}$, calculated
by Kaiser within the HB$\chi$PT \cite{Kaiser2001} to the vector
coupling constant of the neutron $\beta^-$--decay, should be taken
into account for a correct description of the neutron lifetime at the
level of $10^{-5}$.

As has been shown in \cite{Cirigliano2013} the phenomenological
coupling constant $C_P - \bar{C}_P$, introduced at the hadronic level
\cite{Lee1956}--\cite{Severijns2006}, can be related to the effective
coupling constant $\epsilon_P$ of the pseudoscalar interaction of the
{\it up} and {\it down} quarks with left--handed leptonic current by
$C_P - \bar{C}_P = 2 g_P \epsilon_P$, where $g_P = 349(9)$
\cite{Gonzalez-Alonso2014} is the matrix element of the pseudoscalar
quark current caused by strong low--energy interactions. Using the
relation $C_P - \bar{C}_P = 2 g_P \epsilon_P$ \cite{Cirigliano2013} we
have estimated the real and imaginary parts of the phenomenological
coupling constant $C_P - \bar{C}_P$. Having summarized the results,
concerning the constraints on the parameter $\epsilon_P$, obtained in
\cite{Bhattacharya2012, Severijns2019,
  Gonzalez-Alonso2014,Gonzalez-Alonso2016}, and taking into account
that $g_P = 349(9)$ \cite{Gonzalez-Alonso2014}, we have got $ 0
\lesssim {\rm Re}(C_P - \bar{C}_P) \lesssim 0.3$ and ${\rm Im}(C_P -
\bar{C}_P) < 0.2$. Such an estimate agrees well with the analysis of
the contributions of the pseudoscalar interaction BSM to the lifetimes
of charged pions \cite{Severijns2019}.

For the effective coupling constants ${\rm Re}\,C^{(\rm BSM)}_{ps}$
and ${\rm Im}\,C^{(\rm BSM)}_{ps}$, defining the strength of the
contributions of the pseudoscalar interaction BSM to the correlation
coefficients of the electron--energy and angular distribution of the
neutron $\beta^-$--decays, we get $ - 3.5 \times 10^{-5} \lesssim {\rm
  Re}\,C^{(\rm BSM)}_{ps} \lesssim 0$ and ${\rm Im}\,C^{(\rm
  BSM)}_{ps} < - 2.3 \times 10^{-5}$, respectively. This implies that
the effective coupling constant $C^{(\rm BSM)}_{ps}$ is of order
$|C^{(\rm BSM)}_{ps}| \sim 10^{-5}$.

The analysis of contributions of pseudoscalar interactions to the
electron--energy and angular distributions of weak semileptonic decays
of baryons has a long history \cite{Harrington1960}--\cite{Hayen2018}
(see also \cite{Wilkinson1982, Severijns2019}). That is why it is
important to make a comparative analysis of the results obtained in
our work with those in \cite{Wilkinson1982, Severijns2019,
  Harrington1960}--\cite{Hayen2018}.  For the first time the
contributions of pseudoscalar interactions to the correlation
coefficients of electron--energy and angular distributions for weak
semileptonic decays of baryons for polarized parent baryons and
unpolarized decay electrons and baryons were calculated by Harrington
\cite{Harrington1960}. In the notation of Jackson {\it et al.}
\cite{Jackson1957} Harrington calculated the contributions of the
induced pseudoscalar form factor to the Fierz interference term
$b(E_e)$ \cite{Fierz1937} and to the correlation coefficients
$a(E_e)$, $A(E_e)$, $B(E_e)$ and $D(E_e)$, caused by
electron--antineutrino angular correlations and correlations of the
neutron spin with electron and antineutrino 3--momenta,
respectively. The corresponding contributions of pseudoscalar
interactions can be obtained from Eqs.(9) -- (13) of
Ref.\cite{Harrington1960} keeping the leading terms in the large
baryon mass expansion. They read
\begin{eqnarray}\label{eq:B.1}
\hspace{-0.15in}&&\frac{d^5 \delta \lambda_n(E_e, \vec{k}_e,
  \vec{k}_{\bar{\nu}}, \vec{\xi}_n, \vec{\xi}_e)}{dE_e d\Omega_e
  d\Omega_{\bar{\nu}}} \propto -  \frac{{\rm
    Re}(g_1g^*_3)}{|f_1|^2 +
  3|g_2|^2}\,\frac{m^2_e}{M^2}\,\frac{E_{\bar{\nu}}}{E_e} -
 \frac{{\rm Re}(g_1g^*_3)}{|f_1|^2 +
  3|g_2|^2}\,\frac{m^2_e}{M^2}\,\frac{\vec{k}_e\cdot
  \vec{k}_{\bar{\nu}}}{E_eE_{\bar{\nu}}} - \frac{{\rm
    Re}(f_1g^*_3)}{|f_1|^2 +
  3|g_2|^2}\,\frac{m^2_e}{M^2}\,\frac{\vec{\xi}_n\cdot \vec{k}_e}{E_e}
\nonumber\\
\hspace{-0.15in}&&\hspace{1.5in}~~~ - \frac{{\rm Re}(f_1g^*_3)}{|f_1|^2 +
  3|g_2|^2}\,\frac{m^2_e}{M^2}\,\frac{E_{\bar{\nu}}}{E_e}\,\frac{\vec{\xi}_n
  \cdot \vec{k}_{\bar{\nu}}}{E_{\bar{\nu}}} + \frac{{\rm Im}(g_1
  g^*_3)}{|f_1|^2 + 3|g_2|^2}\,\frac{m^2_e}{M^2}\,
\frac{\vec{\xi}_n\cdot \big(\vec{k}_e \times
  \vec{k}_{\bar{\nu}}\big)}{E_e E_{\bar{\nu}}},
\end{eqnarray}
where the first term describes the contribution of pseudoscalar
interactions to the Fierz--like interference term
\cite{Fierz1937}. The analogous corrections can be extracted from the
expressions, calculated by Holstein \cite{Holstein1974} (see Appendix B
of Ref.\cite{Holstein1974}). The corrections of pseudoscalar
interactions to the Fierz--like interference term $\delta b_{ps}(E_e)$
and correlation coefficients $\delta a_{ps}(E_e)$, $\delta
A_{ps}(E_e)$, $\delta B_{ps}(E_e)$ and $\delta D_{ps}(E_e)$,
calculated in Eqs.(\ref{eq:A.5}) and (\ref{eq:A.6}), agree well with
those calculated by Harrington \cite{Harrington1960} (see
Eq.(\ref{eq:B.1})). Since in \cite{Wilkinson1982, Severijns2019,
  Shekhter1960, Bender1968, Armstrong1972, Holstein1974, BB1982,
  Gonzalez-Alonso2014, Hayen2018} the electron--energy and angular
distributions were analyzed for weak semileptonic decays either for
polarized parent baryons and unpolarized decay electrons and baryons
or for unpolarized parent baryons and unpolarized decay electrons and
baryons the overlap of our results with those obtained in
\cite{Wilkinson1982, Shekhter1960, Bender1968, Armstrong1972,
  Holstein1974, BB1982, Gonzalez-Alonso2014, Hayen2018} is at the
level of the corrections shown in Eq.(\ref{eq:B.1}). Indeed, the
contribution of the Fierz--like interference term $\delta b_{ps}(E_e)$
in Eq.(\ref{eq:A.4}) agrees well with the result, obtained by
Wilkinson \cite{Wilkinson1982} and by Gonz\'alez-Alonso and Camalich
\cite{Gonzalez-Alonso2014}
\begin{eqnarray}\label{eq:B.2}
\hspace{-0.3in}&&\frac{d^5 \delta \lambda_n(E_e, \vec{k}_e,
  \vec{k}_{\bar{\nu}}, \vec{\xi}_n, \vec{\xi}_e)}{dE_e d\Omega_e
  d\Omega_{\bar{\nu}}} \propto C'_{ps}\,\lambda\, \frac{E_0 -
  E_e}{E_0} \frac{m_e}{E_e} + \ldots \to - \frac{g_A g_{\rm IP}}{g^2_V +
  3 g^2_A}\,\frac{E_0 - E_e}{M}\,\frac{m_e}{E_e}\nonumber\\
\hspace{-0.3in}&&\hspace{1.5in} - \frac{\lambda}{1 +
  3 \lambda^2}\,g_P{\rm Re}\epsilon_P\,\frac{E_0 -
  E_e}{M}\,\frac{m_e}{E_e} + \ldots,
\end{eqnarray}
where the term proportional to $g_Ag_{\rm I P}$, describing the
contribution of the OPP exchange with $g_{\rm IP} = 2 g_A
M/m^2_{\pi}$, was calculated by Wilkinson (see Table 1 and a
definition of $g_{\rm IP}$ on p.479 of Ref.\cite{Wilkinson1982}),
whereas the second term, caused by the contribution of the
pseudoscalar interaction BSM and where we have taken into account the
relation $C_P - \bar{C}_P = 2 g_P \epsilon_P$ \cite{Cirigliano2013},
was calculated by Gonz\'alez-Alonso and Camalich
\cite{Gonzalez-Alonso2014} (see Eqs.(16) and (17) of
Ref.\cite{Gonzalez-Alonso2014})).

In turn, the contributions of pseudoscalar interactions to the
correlation coefficients, induced by correlations with the electron
spin, were not calculated in \cite{Wilkinson1982, Severijns2019,
  Harrington1960, Shekhter1960, Bender1968, Armstrong1972,
  Holstein1974, BB1982, Gonzalez-Alonso2014, Hayen2018}. Thus, the
calculation of contributions of pseudoscalar interactions to the
correlation coefficients, induced by correlations with the electron
spin, distinguishes our results from those obtained in
\cite{Wilkinson1982, Severijns2019, Harrington1960, Shekhter1960,
  Bender1968, Armstrong1972, Holstein1974, BB1982,
  Gonzalez-Alonso2014, Hayen2018}.  However, we would like to notice
that in the book by Behrens and B\"uhring \cite{BB1982} there is a
capture entitled ``Electron polarization'', concerning an analysis of
a polarization of decay electrons in beta decays. In this capture the
authors propose a most general density matrix, which can be applied to
a description of energy and angular distributions for beta decays by
taking into account a polarization of decay electrons (see Eq.(7.6)
and Eq.(7.7) of Ref.\cite{BB1982}). Of course, by using such a general
density matrix and the technique, developed by Biedenharn and Rose
\cite{Biedenharn1953}, one can, in principle, calculate contributions
of pseudoscalar interactions to the correlation coefficients induced
by correlations with the electron spin. Nevertheless, the calculation
of these corrections were not performed in \cite{BB1982}. The authors
applied such a general density matrix to a calculation of a general
formula for a value of a longitudinal polarization of decay electrons
in beta decays only (see Eq.(7.151) of Ref.\cite{BB1982}). Thus, we
may assert that all corrections of pseudoscalar interactions to the
correlation coefficients, induced by correlations with the electron
spin (see Eq.(\ref{eq:A.5})), and also other terms proportional to the
coupling constants $C'_{ps}$ and $C''_{ps}$ in Eq.(\ref{eq:A.6}) are
new in comparison to the results, obtained in \cite{Wilkinson1982,
  Severijns2019, Harrington1960}--\cite{Hayen2018} and were never
calculated in literature. Moreover, a theoretical accuracy $O(\alpha
E_0/\pi M) \sim 10^{-6}$ and $O(E^2_0/M^2) \sim 10^{-6}$ of the
calculation of a complete set of corrections of order $10^{-3}$
\cite{Ivanov2013, Ivanov2017b, Ivanov2017d, Ivanov2019} including
radiative corrections of order $O(\alpha/\pi)$ and corrections of
order $O(E_0/M)$, caused by the weak magnetism and proton recoil,
makes the contributions of corrections of order $10^{-5}$, induced by
pseudoscalar interactions, observable in principle and important as a
part of theoretical background for experimental searches of
contributions of interactions BSM in asymmetries of the neutron
$\beta^-$--decays with a polarized neutron, a polarized electron and
an unpolarized proton \cite{Abele2016}.

Thus, in this work we have calculated the contributions of
pseudoscalar interactions, induced by the OPP exchange and BSM, to the
complete set of correlation coefficients of the electron--energy and
angular distribution of the neutron $\beta^-$--decays for a polarized
neutron, a polarized electron and an unpolarized proton.  The
corrections to the Fierz interference term $b(E_e)$, the correlation
coefficients $a(E_e)$, $A(E_e)$, $B(E_e)$ and $D(E_e)$, caused by
electron--antineutrino angular correlations and correlations of the
neutron spin with electron and antineutrino 3--momenta, respectively,
and as well as the correlation coefficients, induced by correlations
with the electron spin such as $G(E_e)$, $N(E_e)$ and so on, and also
corrections, given by the terms proportional to the effective coupling
constants $C'_{ps}$ and $C''_{ps}$ in Eq.(\ref{eq:A.6}), are
calculated by using one of the same theoretical technique.  The
agreement of the corrections to the Fierz interference term $b(E_e)$
and the correlation coefficients $a(E_e)$, $A(E_e)$, $B(E_e)$ and
$D(E_e)$ with the results obtained in \cite{Wilkinson1982,
  Severijns2019, Harrington1960}--\cite{Hayen2018} may only confirm a
correctness of our results.

The obtained corrections (see Eq.(\ref{eq:A.5}) and
Eq.(\ref{eq:A.6})), caused by the OPP exchange and the pseudoscalar
interaction BSM, complete the analysis of contributions of
interactions BSM to the correlation coefficients of the neutron
$\beta^-$--decays for a polarized neutron, a polarized electron and an
unpolarized proton carried out in \cite{Ivanov2013, Ivanov2017b,
  Ivanov2017d, Ivanov2019}. For experimental accuracies of about a few
parts of $10^{-5}$ or even better \cite{Abele2016} the exact
analytical expressions of these corrections can be practically
distinguished from the contributions of order $10^{-5}$, caused by the
second class hadronic currents or $G$--odd correlations, calculated by
Gardner and Plaster \cite{Gardner2013} and Ivanov {\it et al.}
\cite{Ivanov2017d, Ivanov2019}.

\section{Acknowledgements}

We thank Hartmut Abele for discussions stimulating the work under
corrections of order $10^{-5}$ to the neutron lifetime and correlation
coefficients of the neutron $\beta^-$--decays for different
polarization states of the neutron and massive decay fermions.  The
work of A. N. Ivanov was supported by the Austrian ``Fonds zur
F\"orderung der Wissenschaftlichen Forschung'' (FWF) under contracts
P31702-N27 and P26636-N20 and ``Deutsche F\"orderungsgemeinschaft''
(DFG) AB 128/5-2. The work of R. H\"ollwieser was supported by the
Deutsche Forschungsgemeinschaft in the SFB/TR 55. The work of
M. Wellenzohn was supported by the MA 23 (FH-Call 16) under the
project ``Photonik - Stiftungsprofessur f\"ur Lehre''. The results
obtained in this paper were reported at International Workshop on
``Current and Future Status of the First-Row CKM Unitarity'', held on
16 - 18 of May 2019 at Amherst Center of Fundamental Interactions,
University of Massachusetts Amherst, USA \cite{CKM2019}.

\newpage

\section*{Appendix A: Calculation of corrections caused by pseudoscalar
  interactions to the electron--energy and angular distribution of the
  neutron $\beta^-$--decays for a polarized neutron, a polarized
  electron and an unpolarized proton }
\renewcommand{\theequation}{A-\arabic{equation}}
\setcounter{equation}{0}

A direct calculation of the corrections, caused by the OPP exchange
and the pseudoscalar interaction BSM \cite{Ivanov2013}, to the
electron--energy and angular distribution of the neutron
$\beta^-$--decays for a polarized neutron, a polarized electron and
an unpolarized proton yields
\begin{eqnarray}\label{eq:A.1}
\hspace{-0.15in}&&\frac{d^5 \delta \lambda_n(E_e, \vec{k}_e,
  \vec{k}_{\bar{\nu}}, \vec{\xi}_n, \vec{\xi}_e)}{dE_e d\Omega_e
  d\Omega_{\bar{\nu}}} = (1 + 3
\lambda^2)\,\frac{G^2_F|V_{ud}|^2}{32\pi^5}\,(E_0 - E_e)^2
\,\sqrt{E^2_e - m^2_e}\, E_e\,F(E_e, Z = 1)\,\frac{1}{E_0
  E_eE_{\bar{\nu}}}\nonumber\\
\hspace{-0.15in}&&\times \,\Big\{C' _{ps}\Big[\lambda\, \Big(- m_e
  (\vec{k}_p\cdot \vec{k}_{\bar{\nu}}) - (\vec{k}_p\cdot
  \vec{k}_e)(\zeta_e\cdot k_{\bar{\nu}}) + (\vec{k}_p\cdot
  \vec{\zeta}_e)(k_e\cdot k_{\bar{\nu}})\Big) + (\vec{\xi}_n \cdot
  \vec{k}_p)\big(m_e E_{\bar{\nu}} + E_e (\zeta_e \cdot k_{\bar{\nu}})
  - \zeta^0_e (k_e\cdot k_{\bar{\nu}})\big)\nonumber\\ 
\hspace{-0.15in}&& + \lambda\,(\vec{\xi}_n \times \vec{k}_p) \cdot
\Big(- E_e(\vec{\zeta}_e \times \vec{k}_{\bar{\nu}}) + \zeta^0_e
(\vec{k}_e\times \vec{k}_{\bar{\nu}}) + E_{\bar{\nu}}(\vec{\zeta}_e
\times \vec{k}_e)\Big) + C''_{ps}\Big[\vec{\zeta}_e\cdot (\vec{k}_e
  \times \vec{k}_{\bar{\nu}})(\vec{\xi}_n\cdot \vec{k}_p) + \lambda
  \Big(E_e\,
\vec{k}_p\cdot (\vec{\zeta}_e \times \vec{k}_{\bar{\nu}})\nonumber\\
\hspace{-0.15in}&& - E_{\bar{\nu}}\, \vec{k}_p\cdot (\vec{\zeta}_e
\times \vec{k}_e) - m_e \,\vec{k}_{\bar{\nu}}\cdot (\vec{\xi}_n \times
\vec{k}_p) - \vec{k}_e \cdot (\vec{\xi}_n \times \vec{k}_p)(\zeta_e
\cdot k_{\bar{\nu}}) + \vec{\zeta}_e\cdot (\vec{\xi}_n \times
\vec{k}_p)(k_e \cdot k_{\bar{\nu}}) \Big)\Big] \Big\}.
\end{eqnarray}
The strength of the contributions of pseudoscalar interactions is
defined by the effective coupling constants $C'_{ps}$ and $C''_{ps}$,
which are the real and imaginary parts of the effective coupling
constant $C_{ps}$ given by
\begin{eqnarray}\label{eq:A.2}
  C_{ps} &=& C^{(\rm OPP)}_{ps} + C^{(\rm
    BSM)}_{ps},\nonumber\\ C^{(\rm OPP)}_{ps} &=& \frac{2\lambda}{1 +
    3\lambda^2}\,\frac{m_e }{m^2_{\pi}}\,E_0 = - 1.47\times
  10^{-5},\nonumber\\
  C^{(\rm BSM)}_{ps} &=& - \frac{1}{1 +
    3\lambda^2}\, \frac{E_0}{2M}\,(C_P - \bar{C}_P) = - 1.17\times
  10^{-4}\,(C_P - \bar{C}_P),\nonumber\\
  C'_{ps} &=& {\rm Re}\,C_{ps}
  = C^{(\rm OPP)}_{ps} + {\rm Re}\, C^{(\rm
    BSM)}_{ps},\nonumber\\ C''_{ps} &=& {\rm Im}\,C_{ps} = {\rm
    Im}\,C^{(\rm BSM)}_{ps}.
\end{eqnarray}
The numerical values are obtained at $\lambda = - 1.27641$, $E_0 =
(m^2_n - m^2_p + m^2_e)/2m_n = 1.2926\,{\rm MeV}$, $m_e = 0.511\,{\rm
  MeV}$ and $M = (m_n + m_p)/2 = 938.918\,{\rm MeV}$ \cite{PDG2018}.
Then, $\zeta_e$ is a 4--polarization vector of the electron
\cite{Itzykson1980}
\begin{eqnarray}\label{eq:A.3}
\zeta_e = (\zeta^0_e, \vec{\zeta}_e) = \Big(\frac{\vec{\xi}_e\cdot
  \vec{k}_e}{m_e}, \vec{\xi}_e + \frac{\vec{k}_e(\vec{\xi}_e\cdot
  \vec{k}_e)}{m_e (E_e + m_e)}\Big)
\end{eqnarray}
obeying the constraints $\zeta^2_e = - \vec{\xi}^{\,2}_e = - 1$ and
$k_e\cdot \zeta_e = 0$. The right-hand-side (r.h.s.) of
Eq.(\ref{eq:A.1}) can be transcribed into the form
\begin{eqnarray}\label{eq:A.4}
\hspace{-0.15in}&&\frac{d^5 \delta \lambda_n(E_e, \vec{k}_e,
  \vec{k}_{\bar{\nu}}, \vec{\xi}_n, \vec{\xi}_e)}{dE_e d\Omega_e
  d\Omega_{\bar{\nu}}} = (1 + 3
\lambda^2)\,\frac{G^2_F|V_{ud}|^2}{32\pi^5}\,(E_0 - E_e)^2
\,\sqrt{E^2_e - m^2_e}\, E_e\,F(E_e, Z =
1)\,\Big\{C'_{ps}\Big[\lambda\,\Big(
  \frac{E_{\bar{\nu}}}{E_0}\,\frac{m_e}{E_e}+
  \frac{m_e}{E_0}\,\frac{\vec{k}_e\cdot \vec{k}_{\bar{\nu}}}{E_e
    E_{\bar{\nu}}}\nonumber\\
\hspace{-0.15in}&& - \frac{m_e}{E_0}\,\frac{\vec{\xi}_e\cdot
  \vec{k}_e}{E_e} - \Big(1 -
\frac{m^2_e}{E_0E_e}\Big)\,\frac{\vec{\xi}_e\cdot
  \vec{k}_{\bar{\nu}}}{E_{\bar{\nu}}} + \Big(1 +
\frac{m_e}{E_0}\Big)\,\frac{(\vec{\xi}_e\cdot
  \vec{k}_e)(\vec{k}_e\cdot \vec{k}_{\bar{\nu}})}{(E_e + m_e) E_e
  E_{\bar{\nu}}}\Big) + \Big(- \frac{m_e}{E_0}\,\frac{\vec{\xi}_n\cdot
  \vec{k}_e}{E_e} -
\frac{m_e}{E_0}\,\frac{E_{\bar{\nu}}}{E_e}\,\frac{\vec{\xi}_n\cdot
  \vec{k}_{\bar{\nu}}}{E_{\bar{\nu}}} + \frac{E_e}{E_0}\,\frac{(\vec{\xi}_n\cdot
  \vec{k}_e)}{E_e}\nonumber\\
\hspace{-0.15in}&&\times \frac{(\vec{\xi}_e\cdot
  \vec{k}_{\bar{\nu}})}{ E_{\bar{\nu}}} + \frac{
  E_{\bar{\nu}}}{E_0}\,\frac{(\vec{\xi}_n\cdot
  \vec{k}_{\bar{\nu}})(\vec{\xi}_e\cdot
  \vec{k}_{\bar{\nu}})}{E^2_{\bar{\nu}}} -
\frac{E_e}{E_0}\,\frac{(\vec{\xi}_n\cdot \vec{k}_e)(\vec{\xi}_e\cdot
  \vec{k}_e)(\vec{k}_e\cdot \vec{k}_{\bar{\nu}})}{(E_e + m_e) E^2_e
  E_{\bar{\nu}}} - \frac{E_{\bar{\nu}}}{E_0}\,\frac{(\vec{\xi}_n\cdot
  \vec{k}_{\bar{\nu}})(\vec{\xi}_e\cdot \vec{k}_e)(\vec{k}_e\cdot
  \vec{k}_{\bar{\nu}})}{(E_e + m_e) E_e E^2_{\bar{\nu}}}\Big) +
\lambda \Big(\frac{E_e E_{\bar{\nu}} - k^2_e}{E_e
  E_0}\nonumber\\
\hspace{-0.15in}&&\times\,(\vec{\xi}_n\cdot \vec{\xi}_e) + \frac{E_e -
  E_{\bar{\nu}}}{E_0}\,\frac{(\vec{\xi}_n\cdot
  \vec{\xi}_e)(\vec{k}_e\cdot \vec{k}_{\bar{\nu}})}{E_e E_{\bar{\nu}}}
+ \frac{E_e - E_{\bar{\nu}} + m_e}{E_0}\,\frac{(\vec{\xi}_n\cdot
  \vec{k}_e)(\vec{\xi}_e\cdot \vec{k}_e)}{(E_e + m_e) E_e} -
\frac{m_e}{E_0}\,\frac{(\vec{\xi}_n\cdot
  \vec{k}_{\bar{\nu}})(\vec{\xi}_e \cdot \vec{k}_e)}{E_e
  E_{\bar{\nu}} } + \frac{E_{\bar{\nu}}}{E_0}\,\frac{(\vec{\xi}_n\cdot
  \vec{k}_e)(\vec{\xi}_e \cdot \vec{k}_{\bar{\nu}})}{E_e
  E_{\bar{\nu}}}\nonumber\\
\hspace{-0.15in}&& -
\frac{E_{\bar{\nu}}}{E_0}\,\frac{(\vec{\xi}_n\cdot
  \vec{k}_{\bar{\nu}})(\vec{\xi}_e \cdot \vec{k}_{\bar{\nu}})}{
  E^2_{\bar{\nu}}} - \frac{E_e}{E_0}\,\frac{(\vec{\xi}_n\cdot
  \vec{k}_e)(\vec{\xi}_e \cdot \vec{k}_e)(\vec{k}_e\cdot
  \vec{k}_{\bar{\nu}})}{(E_e + m_e)E^2_e E_{\bar{\nu}}} +
\frac{E_{\bar{\nu}}}{E_0}\,\frac{(\vec{\xi}_n\cdot
  \vec{k}_{\bar{\nu}})(\vec{\xi}_e \cdot \vec{k}_e)(\vec{k}_e\cdot
  \vec{k}_{\bar{\nu}})}{(E_e + m_e)E_e E^2_{\bar{\nu}}}\Big)\Big] +
C''_{ps}\Big[- \frac{E_e}{E_0}\,\frac{\vec{\xi}_e\cdot (\vec{k}_e \times
    \vec{k}_{\bar{\nu}})(\vec{\xi}_n\cdot \vec{k}_e)}{E^2_e
    E_{\bar{\nu}}}\nonumber\\
\hspace{-0.15in}&& - \frac{E_{\bar{\nu}}}{E_0}\,\frac{\vec{\xi}_e\cdot
  (\vec{k}_e \times \vec{k}_{\bar{\nu}})(\vec{\xi}_n\cdot
  \vec{k}_{\bar{\nu}})}{E_e E^2_{\bar{\nu}}} + \lambda
\Big(\frac{\vec{\xi}_e\cdot(\vec{k}_e \times \vec{k}_{\bar{\nu}})}{E_e
  E_{\bar{\nu}}} + \frac{m_e}{E_0}\,\frac{\vec{\xi}_n\cdot (\vec{k}_e
  \times \vec{k}_{\bar{\nu}})}{E_e E_{\bar{\nu}}} +
\frac{E_e}{E_0}\,\frac{\vec{\xi}_n\cdot (\vec{\xi}_e \times
  \vec{k}_e)}{E_e} + \frac{E_{\bar{\nu}}}{E_0}\,
\frac{\vec{\xi}_n\cdot (\vec{\xi}_e \times
  \vec{k}_{\bar{\nu}})}{E_{\bar{\nu}}}\nonumber\\
\hspace{-0.15in}&& - \frac{E_e}{E_0}\,
\frac{\vec{\xi}_n\cdot(\vec{\xi}_e \times \vec{k}_e)(\vec{k}_e \cdot
  \vec{k}_{\bar{\nu}})}{E^2_e E_{\bar{\nu}}} -
\frac{E_{\bar{\nu}}}{E_0}\, \frac{\vec{\xi}_n\cdot (\vec{\xi}_e \times
  \vec{k}_{\bar{\nu}})(\vec{k}_e \cdot \vec{k}_{\bar{\nu}})}{ E_e
  E^2_{\bar{\nu}}} + \frac{E_{\bar{\nu}}}{E_0}\,
\frac{\vec{\xi}_n\cdot (\vec{k}_e \times
  \vec{k}_{\bar{\nu}})(\vec{\xi}_e \cdot \vec{k}_{\bar{\nu}})}{ E_e
  E^2_{\bar{\nu}}}- \frac{E_{\bar{\nu}}}{E_0}\,\frac{\vec{\xi}_n\cdot
  (\vec{k}_e \times \vec{k}_{\bar{\nu}})(\vec{\xi}_e \cdot
  \vec{k}_e)}{(E_e + m_e) E_e E_{\bar{\nu}}}\Big) \Big] \Big\}.
\end{eqnarray}
We obtain the following contributions to the correlation coefficients
\begin{eqnarray*}
  \hspace{-0.15in}\delta
    \zeta_{ps}(E_e) &=&0\;,\;\delta
    b_{ps}(E_e) = C'_{ps}\,\lambda\,\frac{E_0 -
    E_e}{E_0}\;,\;\delta a_{ps}(E_e) = C'_{ps}\,
  \lambda\, \frac{m_e}{E_0}\;,\;\delta A_{ps}(E_e) =
  -\,C'_{ps}\, \frac{m_e}{E_0},\nonumber\\
 \end{eqnarray*}
  \begin{eqnarray}\label{eq:A.5}  
   \hspace{-0.15in}\delta B_{ps}(E_e) &=& -\,C'_{ps}\,
   \frac{m_e}{E_0}\,\frac{E_0 - E_e}{E_e}\;,\;\delta K_{n
       ps}(E_e) =\delta Q_{n
       ps}(E_e) = 0\;,\; \delta
   G_{ps}(E_e) = - C'_{ps}\,\lambda\,\frac{m_e}{E_0},\nonumber\\ 
   \hspace{-0.15in}\delta H_{ps}(E_e) &=& -\,C'_{ps}\,\lambda\, \Big(1
   - \frac{m^2_e}{E_0 E_e}\Big)\;,\;\delta Q_{e ps}(E_e) =
   C'_{ps}\,\Big(\lambda \,\frac{2 E_e - E_0 + m_e}{E_0} + (\lambda -
   1)\,\frac{1}{3}\frac{E_0 - E_e}{E_0}\Big),\nonumber\\
   \hspace{-0.15in}\delta K_{eps}(E_e)&=& C'_{ps}\,\lambda\,\Big(1 +
   \frac{m_e}{E_0}\Big)\;,\; \delta N_{ps}(E_e) =
   C'_{ps}\Big(\lambda\,\frac{- 2 E^2_e + E_0 E_e + m^2_e}{E_0 E_e} +
   (1 - \lambda)\,\frac{1}{3}\,\frac{E_0 - E_e}{E_0}\Big), \nonumber\\
   \hspace{-0.15in}\delta D_{ps}(E_e)&=&
   C''_{ps}\,\lambda\,\frac{m_e}{E_0}\;,\; \delta R_{ps}(E_e) =
   C''_{ps}\,\Big( - \lambda\,\frac{E_e}{E_0} + (1 + 2
   \lambda)\,\frac{1}{3}\,\frac{E_0 - E_e}{E_0}\Big)\;,\; \delta
   L_{ps}(E_e) = C''_{ps}\,\lambda.
\end{eqnarray}
In terms of corrections to the correlation coefficients
Eq.(\ref{eq:A.5}) the correction to the electron--energy and angular
distribution Eq.(\ref{eq:A.4}) is given by
\begin{eqnarray}\label{eq:A.6}
\hspace{-0.15in}&&\frac{d^5 \delta \lambda_n(E_e, \vec{k}_e,
  \vec{k}_{\bar{\nu}}, \vec{\xi}_n, \vec{\xi}_e)}{dE_e d\Omega_e
  d\Omega_{\bar{\nu}}} = (1 + 3
\lambda^2)\,\frac{G^2_F|V_{ud}|^2}{32\pi^5}\,(E_0 - E_e)^2
\,\sqrt{E^2_e - m^2_e}\, E_e\,F(E_e, Z = 1)\,\Big\{ \delta
b_{ps}(E_e)\,\frac{m_e}{E_e} + \delta
a_{ps}(E_e)\,\frac{\vec{k}_e\cdot \vec{k}_{\bar{\nu}}}{E_e
  E_{\bar{\nu}}}\nonumber\\
\hspace{-0.15in}&&  + \delta
A_{ps}(E_e)\,\frac{\vec{\xi}_n\cdot \vec{k}_e}{E_e} + \delta
B_{ps}(E_e)\, \frac{\vec{\xi}_n\cdot
  \vec{k}_{\bar{\nu}}}{E_{\bar{\nu}}} + \delta G_{ps}(E_e)\,\frac{\vec{\xi}_e
  \cdot \vec{k}_e}{E_e} + \delta H_{ps}(E_e)\,\frac{\vec{\xi}_e \cdot
  \vec{k}_{\bar{\nu}}}{E_{\bar{\nu}}} + \delta Q_{e
  ps}(E_e)\,\frac{(\vec{\xi}_n\cdot \vec{k}_e)( \vec{k}_e\cdot
  \vec{\xi}_e)}{(E_e + m_e) E_e} + \delta K_{e ps}(E_e)\nonumber\\
\hspace{-0.15in}&&\times\,\frac{(\vec{\xi}_e\cdot \vec{k}_e)(
  \vec{k}_e\cdot \vec{k}_{\bar{\nu}})}{(E_e + m_e)E_e E_{\bar{\nu}}} +
\delta N_{ps}(E_e) \,(\vec{\xi}_n\cdot \vec{\xi}_e) + \delta
R_{ps}\,\frac{\vec{\xi}_n\cdot (\vec{k}_e \times \vec{\xi}_e)}{E_e} +
\delta L_{ps}\,\frac{\vec{\xi}_n\cdot (\vec{k}_e \times
  \vec{k}_{\bar{\nu}})}{E_e E_{\bar{\nu}}} +
C'_{ps}\Big[\lambda\,\frac{E_0 - E_e}{E_0}\frac{(\vec{\xi}_n\cdot
    \vec{\xi}_e)( \vec{k}_e\cdot \vec{k}_{\bar{\nu}})}{E_e
    E_{\bar{\nu}}}\nonumber\\
\hspace{-0.15in}&& + \frac{(1 - \lambda) E_e + \lambda E_0
}{E_0}\frac{(\vec{\xi}_n\cdot \vec{k}_e)( \vec{\xi}_e\cdot
  \vec{k}_{\bar{\nu}})}{E_e E_{\bar{\nu}}} -
\lambda\,\frac{m_e}{E_0}\,\frac{(\vec{\xi}_n\cdot
  \vec{k}_{\bar{\nu}})(\vec{\xi}_e\cdot \vec{k}_e)}{E_e E_{\bar{\nu}}}
- (1 + \lambda) \frac{E_e}{E_0}\,\frac{(\vec{\xi}_n\cdot
  \vec{k}_e)(\vec{\xi}_e \cdot \vec{k}_e)(\vec{k}_e\cdot
  \vec{k}_{\bar{\nu}})}{(E_e + m_e)E^2_e E_{\bar{\nu}}} + (1 -
\lambda)\,\frac{E_0 - E_e}{E_0}\nonumber\\
\hspace{-0.15in}&&\times \,\Big(\frac{(\vec{\xi}_n\cdot
  \vec{k}_{\bar{\nu}})(\vec{\xi}_e\cdot
  \vec{k}_{\bar{\nu}})}{E^2_{\bar{\nu}}} -
\frac{1}{3}\,\vec{\xi}_n\cdot \vec{\xi}_e\Big) + (\lambda -
1)\,\frac{E_0 - E_e}{E_0}\,\Big(\frac{(\vec{\xi}_n\cdot
  \vec{k}_{\bar{\nu}})(\vec{k}_e\cdot
  \vec{k}_{\bar{\nu}})}{E^2_{\bar{\nu}}} -
\frac{1}{3}\,\vec{\xi}_n\cdot \vec{k}_e \Big) \frac{(\vec{\xi}_e\cdot
  \vec{k}_e)}{(E_e + m_e) E_e}\Big] + C''_{ps}\Big[- \frac{E_e}{E_0}
  \nonumber\\  
\hspace{-0.15in}&&\times \,\frac{\vec{\xi}_e\cdot (\vec{k}_e \times
  \vec{k}_{\bar{\nu}})(\vec{\xi}_n\cdot \vec{k}_e)}{E^2_e
  E_{\bar{\nu}}} + \lambda\,\frac{E_0 - E_e}{E_0}
\,\frac{\vec{\xi}_n\cdot (\vec{\xi}_e \times
  \vec{k}_{\bar{\nu}})}{E_{\bar{\nu}}} -
\lambda\,\frac{E_e}{E_0}\,\frac{\vec{\xi}_n\cdot (\vec{\xi}_e \times
  \vec{k}_e)(\vec{k}_e\cdot \vec{k}_{\bar{\nu}})}{E^2_e E_{\bar{\nu}}}
- \lambda\,\frac{E_0 - E_e}{E_0} \frac{\vec{\xi}_n\cdot (\vec{k}_e
  \times \vec{k}_{\bar{\nu}})(\vec{\xi}_e\cdot \vec{k}_e)}{(E_e + m_e)
  E_e E_{\bar{\nu}}} \nonumber\\
\hspace{-0.15in}&& - \frac{E_0 - E_e}{E_0} \Big(\frac{\vec{\xi}_e\cdot
  (\vec{k}_e \times \vec{k}_{\bar{\nu}})(\vec{\xi}_n\cdot
  \vec{k}_{\bar{\nu}})}{E_e E^2_{\bar{\nu}}} - \frac{1}{3}
\frac{\vec{\xi}_e\cdot (\vec{k}_e \times \vec{\xi}_n)}{E_e} \Big) -
\lambda\,\frac{E_0 - E_e}{E_0} \Big(\frac{\vec{\xi}_n\cdot
  (\vec{\xi}_e \times \vec{k}_{\bar{\nu}})(\vec{k}_e\cdot
  \vec{k}_{\bar{\nu}})}{E_e E^2_{\bar{\nu}}} - \frac{1}{3}
\frac{\vec{\xi}_n\cdot (\vec{\xi}_e \times \vec{k}_e)}{E_e} \Big)
\nonumber\\
 \hspace{-0.15in}&&+ \lambda\,\frac{E_0 - E_e}{E_0}\Big(
 \frac{\vec{\xi}_n\cdot (\vec{k}_e \times
   \vec{k}_{\bar{\nu}})(\vec{\xi}_e \cdot \vec{k}_{\bar{\nu}})}{ E_e
   E^2_{\bar{\nu}}} - \frac{1}{3}\,\frac{\vec{\xi}_n\cdot (\vec{k}_e
   \times \vec{\xi}_e)}{E_e}\Big)\Big]\Big\}.
\end{eqnarray}
This correction to the electron--energy and angular distribution
together with the results obtained in \cite{Gudkov2006, Ivanov2013,
  Ivanov2017b, Ivanov2017d, Ivanov2019}, can be used for experimental
analyses of asymmetries and correlation coefficients of the neutron
$\beta^-$--decays for a polarized neutron, a polarized electron and
an unpolarised proton with experimental uncertainties of a few parts
of $10^{-5}$ \cite{Abele2016}.

\end{document}